\documentclass[Manuscript,12pt]{elsarticle}

\usepackage{graphicx}
\usepackage{amsmath}
\usepackage{amssymb}
\usepackage{amsthm}
\usepackage{relsize}
\usepackage{multirow}
\usepackage{rotating}
\usepackage{bm}
\usepackage{url}
\usepackage{gensymb}
\usepackage{calc}
\usepackage{multicol}
\usepackage{color}
\usepackage[dvipsnames]{xcolor}
\usepackage{epstopdf}
\usepackage{svg}
\usepackage{subeqnarray}
\usepackage[all]{xy}
\usepackage{array}
\usepackage{ctable}
\usepackage{tabu}
\usepackage{longtable}
\usepackage{caption}
\usepackage{dashrule}
\usepackage{stmaryrd}
\usepackage[version=3]{mhchem}

\journal{Combustion and flame}

\begin{document}

\begin{frontmatter}


\title{A Multipurpose Reduced Mechanism for Ethanol Combustion}



\author[label1]{Alejandro Mill\'an-Merino\corref{cor1} }
\author[label1]{Eduardo Fern\'andez-Tarrazo }
\author[label1]{Mario S\'anchez-Sanz}
\author[label2]{Forman A. Williams }

\address[label1]{Dept. Ingenier\'ia T\'ermica y de Fluidos,
	Universidad Carlos III de Madrid, Legan\'es 28911, Spain}
\address[label2]{Dept. Mechanical and Aerospace Engineering, University of California San Diego, La Jolla, USA}

\cortext[cor1]{almillan@ing.uc3m.es}
\begin{abstract}
\noindent
New multipurpose skeletal and reduced chemical-kinetic mechanisms for ethanol combustion are developed, along the same philosophical lines followed in our previous work on methanol. The resulting skeletal mechanism contains 66 reactions, only 19 of which are reversible, among 31 species, and the associated reduced mechanism contains 14 overall reactions among 16 species, obtained from the skeletal mechanism by placing \ce{CH3CHOH}, \ce{CH2CH2OH}, \ce{CH3CO}, \ce{CH2CHO}, \ce{CH2CO}, \ce{C2H3}, \ce{C2H5}, \ce{C2H6}, \ce{S-CH2}, \ce{T-CH2}, \ce{CH4}, \ce{CH2OH}, \ce{CH3O}, \ce{HCO}, and \ce{O} in steady state. For the reduced mechanism, the steady-state relations and rate expressions are arranged so that computations can be made sequentially without iteration. Comparison with experimental results for autoignition, laminar burning velocities, and counterflow flame structure and extinction, including comparisons with the 257-step, 54-species detailed San Diego Mechanism and five other mechanisms in the literature, support the utility of the skeletal and reduced mechanisms, showing, for example, that, in comparison with the San Diego mechanism, they decrease the computational cost by 70\% and 80\%, respectively. Measures of computation times and of extents of departures from experimental values are defined and employed in evaluating results. Besides contributing to improvements in understanding of the mechanisms, the derived simplifications may prove useful in a variety of computational studies. 
\end{abstract}

\begin{keyword}
ethanol combustion \sep reduced chemistry \sep Multipurpose kinetics


\end{keyword}

\end{frontmatter}


\section{Introduction}
\noindent
Because of concerns about greenhouse gas emissions (essentially CO$_2$), the use of biofuels has been growing steadily in the last decade \cite{Gold17}. In countries like the United Stated, Brazil, India, and China, the biofuel used most extensively is ethanol.
The characteristics of many biofuels such as ethanol make them suitable for applications in existing propulsion and energy-generation devices, as an additive to gasoline, or, in some cases, as stand-alone fuels themselves. The lower energy density of biofuels is offset partially in many applications by an increase of the octane number of the resulting fuel, as occurs in gasoline-ethanol mixtures, for example, \cite{RFA16}, thereby permitting higher compression ratios to increase the engine efficiency. Although questions remain concerning their overall future promise and the influences of biofuels on emissions of toxic pollutants, there clearly is sufficient interest in ethanol to motivate studies of the chemical kinetics of its combustion. Numerical fuel-specific simulations of combustion in bio-fueled engines, which play ever-increasing roles in studies of design optimizations to increase efficiencies, for example, benefit greatly from the availability of chemical-kinetic mechanisms that are as short and as accurate as possible, the development of which for ethanol is the objective of the present work.\\
\noindent
The specific aim of the present work is to derive multipurpose reduced ethanol kinetic mechanisms that are able to describe, accurately and efficiently, both premixed flame propagation and autoignition, as well as structures of diffusion flames and flame extinction, for a range of application-relevant conditions of pressure, temperature, and equivalence ratio. These laminar combustion processes span a sufficient set of phenomena that, once they all are encompassed, the same chemistry is likely to work well for describing practical turbulent combustion in engines, which generally will involve a combination of these processes. The aim here parallels that of our previous study of methanol \cite{FSSW2016}. Since the length and complexity of the mechanisms necessarily increase as the number of carbon atoms in the alcohol increases, the present task is more difficult. The objective nevertheless is found here to be achievable, and it may also be possible to address even higher alcohols in the same vein in the future.\\
\section{Quantitative selection criteria}
\noindent
Besides general qualitative criteria, it is worthwhile to evaluate quantitative criteria for comparisons of selections. Two such types of criteria are introduced here. One is the CPU time; the CPU time for autoignition is defined here as the time needed to complete the simulation starting with the prescribed initial conditions, but for other problems, such as premixed flames, it is strongly dependent on exactly how the problem is addressed. For premixed flames, in particular, in order to establish a neutral test, each mechanism was run starting from a converged solution at a pressure slightly above the pressure of the atmosphere being investigated, and the time needed to convergence from that condition was measured. Although measures for other problems were not computed, they may be expected to lie in the same relative order as those for premixed flames.\\
\noindent
The second type of quantitative criterion considered was a normalized deviation from experimental data points, defined as a percentage for any property $Q$ according to
\begin{equation}\label{eq:deltaQ}
\Delta_Q = \frac{100}{N} \sum_{i=1}^{N} \frac{|Q_{\rm sim,i}-Q_{\rm exp,i}|}{Q_{\rm exp,i}},
\end{equation}
where $N$ is the number of experimental data points, and the subscripts "sim" and "exp" refer to the numerically computed result and the experimentally measured value, respectively. This last choice, one of many possibilities, does not factor in any possible experimental error and therefore could be misleading. To do better, however, would require unavailable knowledge of experimental errors. The two quantitative criteria, taken together, should provide a reasonable indication of relative performance.
\section{Selection of a detailed mechanism}
\begin{figure}[!htp]
	\centering
	\begin{tabular}{cc}
		\includegraphics[width=.47\textwidth]{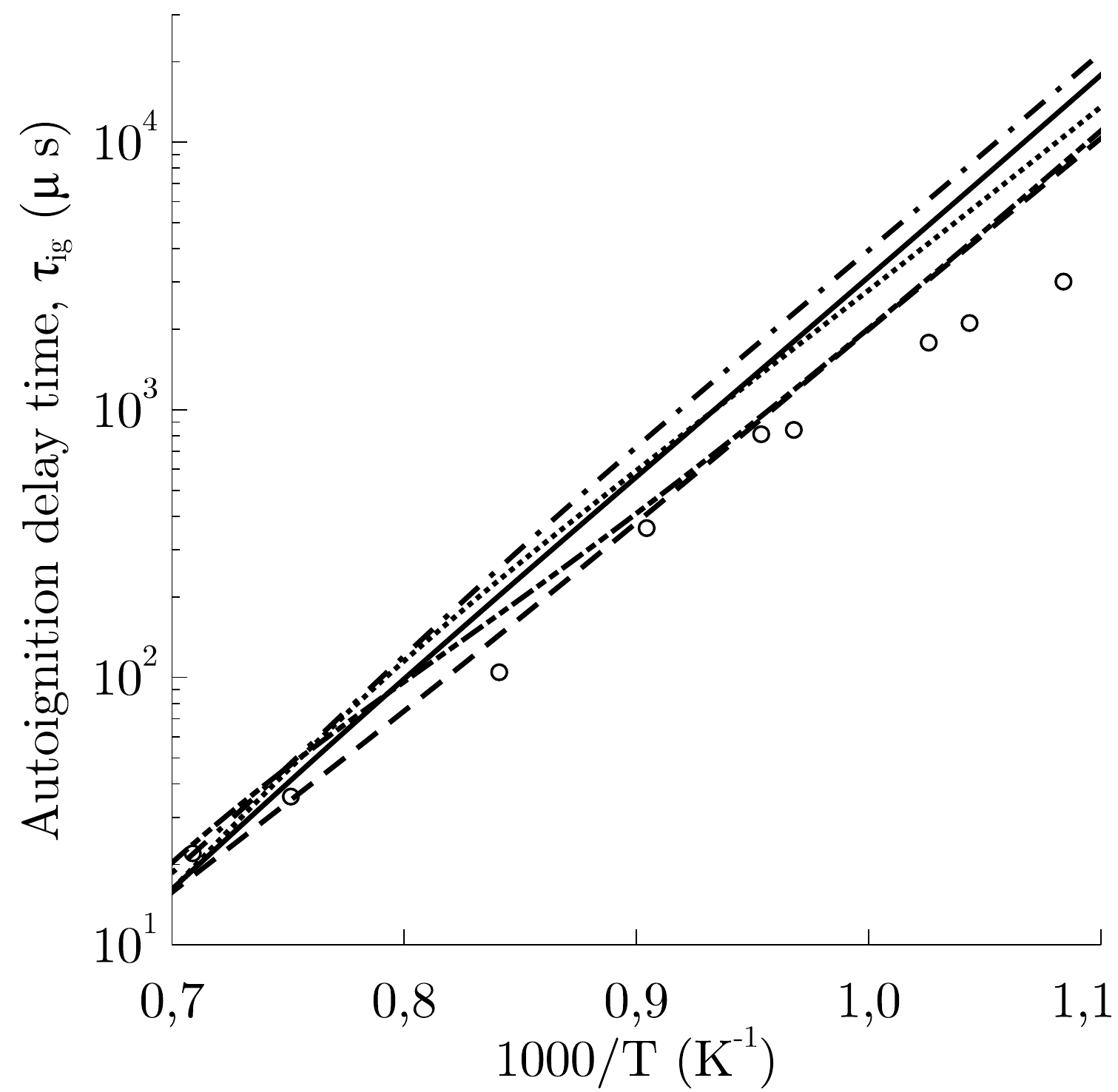} &
		\includegraphics[width=.47\textwidth]{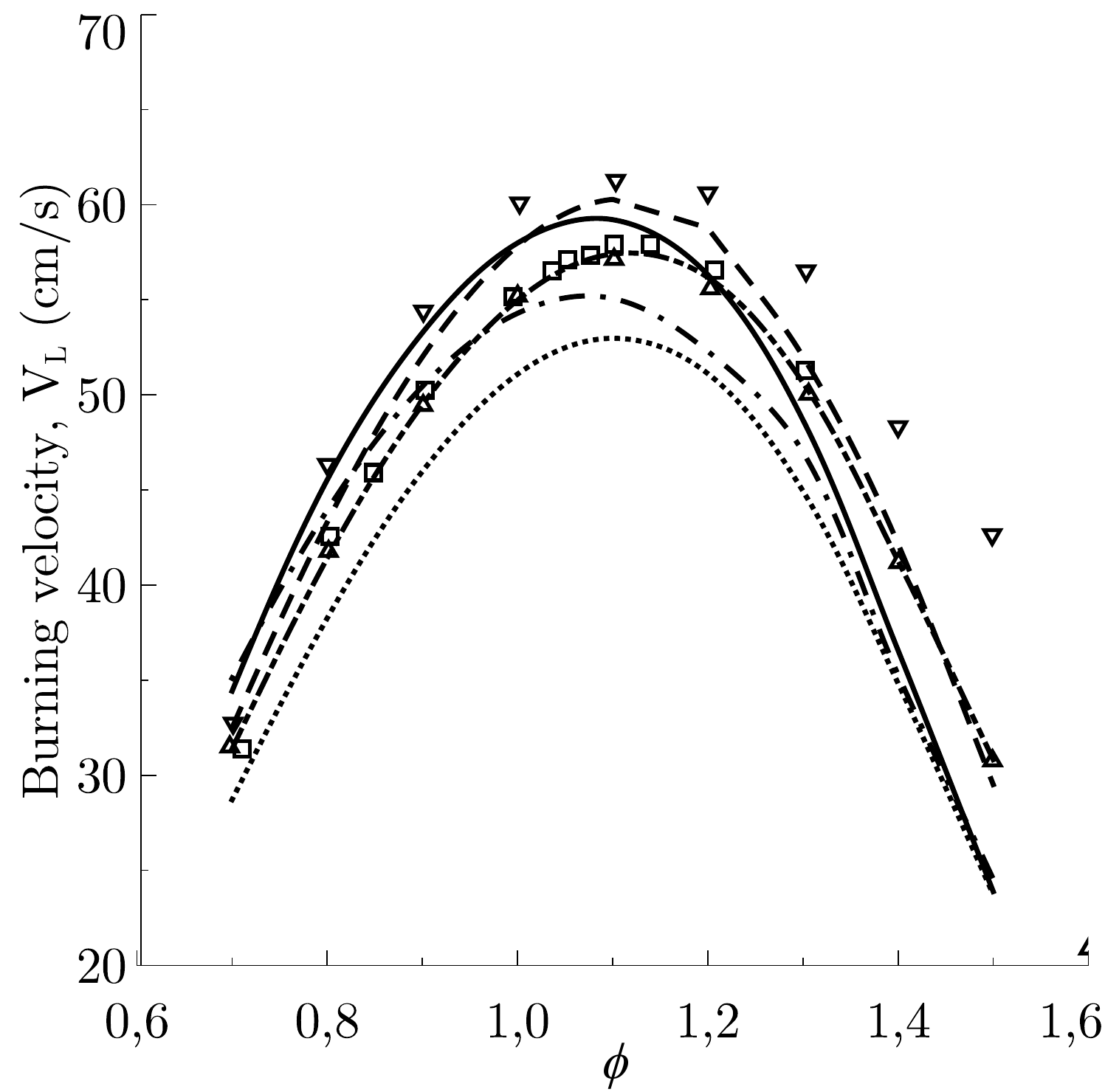} \\
		(a) & (b) \\
	\end{tabular}  
	\caption{
		Performance of the detailed mechanisms under analysis. Subfigure \emph{a} shows the ignition times of a stoichiometric ethanol-air mixture as a function of temperature, for a pressure $p=13$ bar. Subfigure \emph{b} show the flame propagation velocity $V_L$ in a mixture of ethanol-air as a function of the equivalence ratio $\phi$, at atmospheric pressure and fresh-mixture temperature $T=373$K. The lines represent the results obtained with detailed mechanisms and the symbols correspond to experimental results, as follows:\\ 
		{\color{black} \hdashrule[0.5ex][c]{10mm}{2pt}{5mm 0mm} }: San Diego Mechanism
		, {\color{black} \hdashrule[0.5ex][c]{10mm}{2pt}{3mm 3pt} }: Aramcomech 2.0
		, {\color{black} \hdashrule[0.5ex][c]{12mm}{2pt}{3mm 3pt 1pt 3pt} }: Lawrence Livermore
		, {\color{black} \hdashrule[0.5ex][c]{12mm}{2pt}{3mm 2pt 2pt 1pt 2pt 1pt} }: CRECK\\
		, {\color{black} \hdashrule[0.5ex][c]{15mm}{2pt}{2pt 2pt} }: ELTE
		, {\color{black} $\circ$}: Heufer and Olivier \cite{Heu10}; other autoignition data (not shown) gave similar results \cite{Heu10, Noo10, Nat82, Dun91}
		, {\color{black} $\square$}: Aghsaee et al. \cite{Aghsaee2015}
		, {\color{black} $\triangle$}: Knorsch et al.\cite{Kno14}
		, {\color{black} $\triangledown$}: Varea et al. \cite{Var13}   }
	\label{fig:detailed_chemistry}  
\end{figure}

\noindent
The first task in a systematic reduction process is to select a detailed mechanism as a starting point for the reduction. Ethanol is well studied, so that more than a dozen detailed mechanisms are available for it in the literature. The selection is made here on the basis of the computation cost and the degree of agreement with experimental data, as indicated above. The mechanisms chosen for comparison here are the San Diego mechanism \cite{SDa} (SD), the AramcoMech v2.0 \cite{ARAMCO} (ARAMCO), the C$_1$-C$_3$ mechanism from CRECK Modelling group \cite{CRECK} (CRECK), the Lawrence Livermore National Laboratory mechanism \cite{LLNL} (LLNL) and the mechanism developed by the Laboratory for Chemical Kinetics E\"otv\"os University \cite{ELTE} (ELTE). The publication of the last of these mechanisms, an optimization study that began with the San Diego mechanism as a starting trial, may be consulted for a listing of other ethanol mechanisms. The five selected here are among the most successful found in that study. \\  
\noindent
Table 1 lists the sizes of these mechanisms as well as the error estimates and CPU times for each, for the logarithm of the autoignition time and for the laminar burning velocity. It is seen that, although the accuracies of the five detailed mechanisms are comparable, there are significant differences in the computational costs. The San Diego and ELTE mechanisms are seen from this list to be the most attractive ones from the viewpoint of size and CPU time. Since they differ little in that respect, the San Diego mechanism is selected here.\\
\noindent
As a further comparison, for the five mechanisms Fig. \ref{fig:detailed_chemistry} shows autoignition delays for stoichiometric ethanol-air mixtures at 13 bar as functions of temperature, as well as laminar burning velocities at normal atmospheric pressure and at an initial temperature of 373K (just above the dew point) as functions of the equivalence ratio. There are many other measurements of burning velocities \cite{Aghsaee2015}-\cite{Lia14}, 
but since the resulting comparisons are similar, they are not shown here. The computations performed to generate the figures employed Cosilab \cite{Cos}, which uses a mixture-average model for the transport properties that are needed in the flame calculations. Cosilab \cite{Cos} also was used to obtain computational results reported later, the results being checked by additional computations with CHEMKIN \cite{Chem}, {exercising its multicomponent transport option}. The differences between the predictions of the five mechanisms are seen in this figure to be comparable with differences in experimental values, with the ELTE mechanism producing slightly lower burning velocities under the conditions shown, a result that also extends to the burning-velocity dependence on pressure and initial temperature (not shown). All of the available mechanisms have a tendency to overpredict measured autoignition times at the lower temperatures, a deficiency that deserves further investigation and is not likely to be related to cool-flame types of chemistry, since ethanol should not exhibit that kind of behavior; the differences may be related to well-known shock-tube difficulties encountered at lower temperatures. 
\begin{table}[!htp]
	\centering
	\begin{tabular}{|c|cc|cc|cc|} 
		\hline 
		& & & & & & \\[-0.2cm]
		\multirow{3}{*}{Mechanism} &\multicolumn{2}{c|}{\multirow{2}{*}{Size}}& \multicolumn{2}{c|}{Error $(\%)$ }  & \multicolumn{2}{c|}{\multirow{2}{*}{Average CPU time}}  \\[0.1cm]
		&\multicolumn{2}{c|}{} &\multicolumn{2}{c|}{defined by (\ref{eq:deltaQ})}  & \multicolumn{2}{c|}{}  \\[0.1cm]
		& Reactions 
		& Species
		&$\Delta_{\log(\tau_{ig})}$
		& $\Delta_{V_L}$
		&$t_{\tau_{ig}}$(s)
		&$t_{V_L}$(min)
		\\[0.1cm]
		\hline 
		& & & & & & \\[-0.15cm]
		SD          &235   &47    &5.21    &7.51            &$<$1      &11              \\[0.1cm]
		ARAMCO      &2716  &497   &7.53    &6.49            &101     & 6999         \\[0.1cm]
		LLNL        &383   &54    &8.43    &8.25            &1       &13              \\[0.1cm]
		CRECK       &2642  &107   &6.82    &6.56            &18      &116             \\[0.1cm]
		ELTE        &251   &49    &6.57    &12.51           &$<$1      &7              \\[0.1cm]
		\hline
		& & & & & &  \\[-0.1cm]
		Skeletal    &66    &31    &5.92  &7.11             &$<$1       &3  \\[0.1cm]
		Reduced     &14    &16    &6.39  &6.37             &$<$1       &2  
		\\[0.2cm] 
		\hline
	\end{tabular}\\
	\caption{ Global results and computational time associated for a each mechanism.}
	\label{tab:01}
\end{table}
\section{The skeletal mechanism}
\noindent
In order to reduce the computational cost associated with the chemistry, efforts were focused on finding a minimal set of species and reactions capable of reproducing autoignition times and premixed-flame propagation velocities of ethanol-air mixtures over a wide range of equivalence ratios, ambient pressure and temperature. 
These problems were considered to be the most relevant for the applications and sufficiently generic for one to expect that the chemistry so developed may well be correct for other applications as well.
In deriving the skeletal mechanism, the complete detailed description was used as a starting point followed by proceeding to discarding chemical species and elementary reactions with negligible influences, to achieve sufficiently accurate predictions of the relevant combustion phenomena.  
The final (minimal) skeletal mechanism, consists of the 31 species and 66 reactions shown as table \ref{table:skeletal}, in appendix A. 
Most of the reactions in the table are indicated to be irreversible; the backward rates are retained only in a few of the reactions, as indicated, because their effects were found to be important for achieving the desired accuracy.  In all of the comparisons performed (like those in Fig, 1), it was found to be practically impossible to distinguish by eye the differences between the predictions of the detailed and skeletal mechanisms. For that reason there is no point in exhibiting here any figures comparing the two.\\
\noindent
\noindent
Saxena and Williams \cite{Sax07} presented and discussed a reaction-path diagram, based on detailed chemistry, for a partially premixed ethanol flame. A somewhat different approach was applied here, investigating the skeletal mechanism for both lean and rich premixed flames; differences from the previous procedure are that the final formation of \ce{CO} and \ce{CO2} is included in the diagram here, and the percentages shown are based on the carbon atom flux rather than on the attacking species. The results appear on the left in Fig. \ref{fig:pathway_skeletal}; in those subfigures, dashed circles identify steady-state species. The results exhibit general qualitative agreement with pathways given by Saxena and Williams \cite{Sax07}, even though the conditions are not exactly the same. In both lean and rich flames, 
ethanol oxidation to \ce{CO2} begins essentially along three abstraction routes \cite{Sarathy2014}, producing ethylene through $\beta$-hydroxyethyl (\ce{CH2CH2OH}, on the left), and leading to methyl through both $\alpha$-hydroxyethyl ({\ce{CH3CHOH}}, in the center) and ethoxy ({\ce{CH3CH2O}}, on the right). In addition, while the partially premixed flame (with an adiabatic flame temperature of 2240 K) exhibited more than 20 percent of the initial fuel removal occurring through unimolecular decomposition to ethylene and water, that path is negligible for the present lean flame but of comparable importance in the rich case. 
All routes lead, of course, to an appreciable amount of $\ce{CO2}$, mainly through $\ce{CO}$,  although for rich flames most of the carbon remains in CO, and 3\% bypasses CO to form CO2 directly along the ethylene route, and where, moreover, as much as 8 percent of the carbon remains in compounds containing two carbon atoms, primarily acetaldehyde. The difference in the final conversion ratio from $\ce{CO}$ to $\ce{CO2}$ between the lean and the rich flames is captured well by the skeletal mechanism.
As is well known, acetaldehyde, obtained through the $\alpha$-hydroxyethyl and ethoxy paths, plays an important role in the oxidation, being consumed by H-atom abstractions to produce the radicals 
$\ce{CH2CHO}$ and $\ce{CH3CO}$, as well as by C-C bond breaking to produce $\ce{CH3}$ and $\ce{HCO}$ directly, for rich flames. 
The ethylene oxidation route becomes dominant at high equivalence ratios, while the most important route for lean flames appears to be the ethoxy route. Both routes are interconnected by many intermediate species generated in the oxidation process, most of them being in steady state.
\section{The systematically reduced mechanism} \label{sec:reduced}
\begin{figure}[!htp]
	\centering
	\begin{tabular}{cc}
		\includegraphics[width=.38\textwidth]{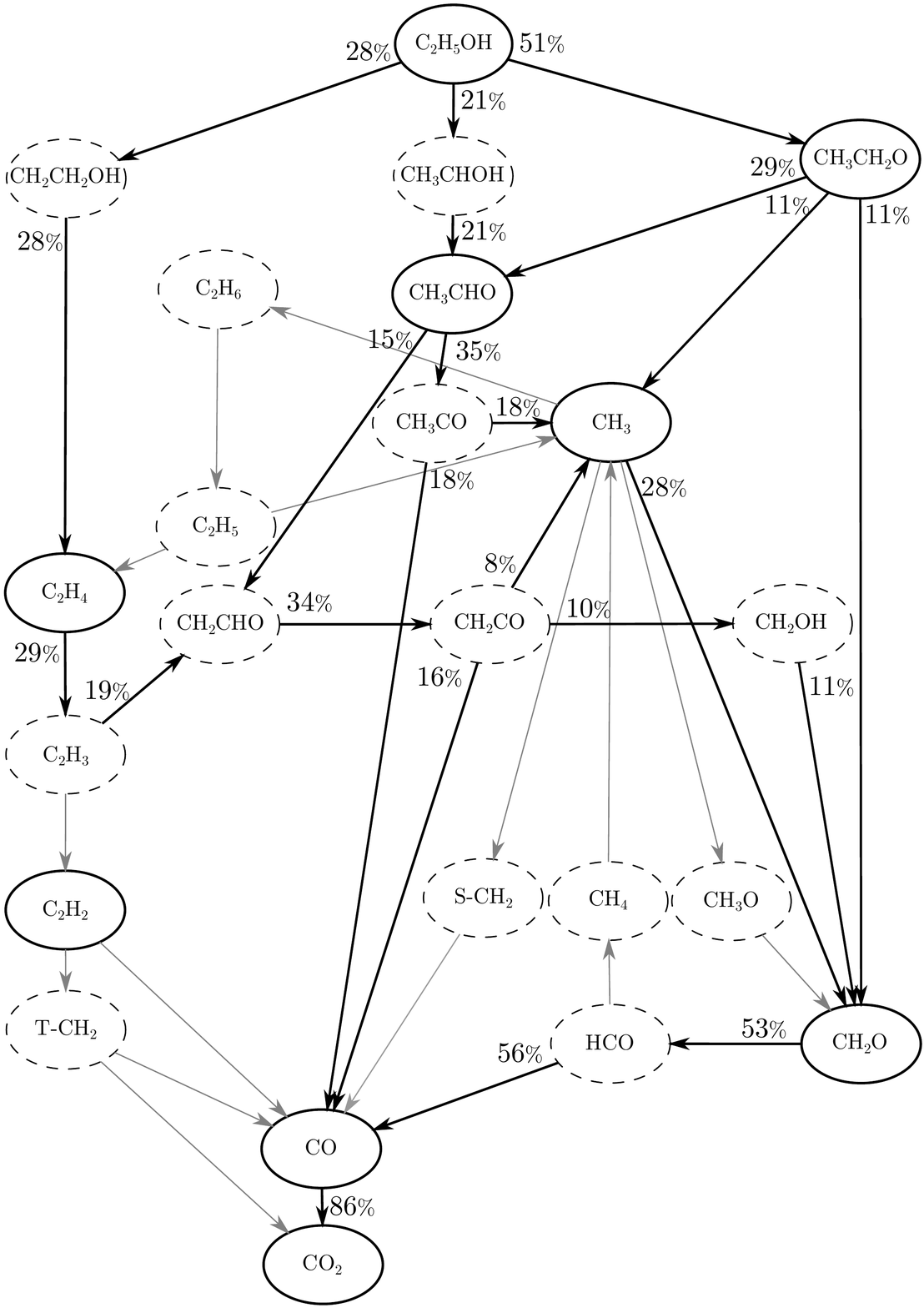} &
		\includegraphics[width=.38\textwidth]{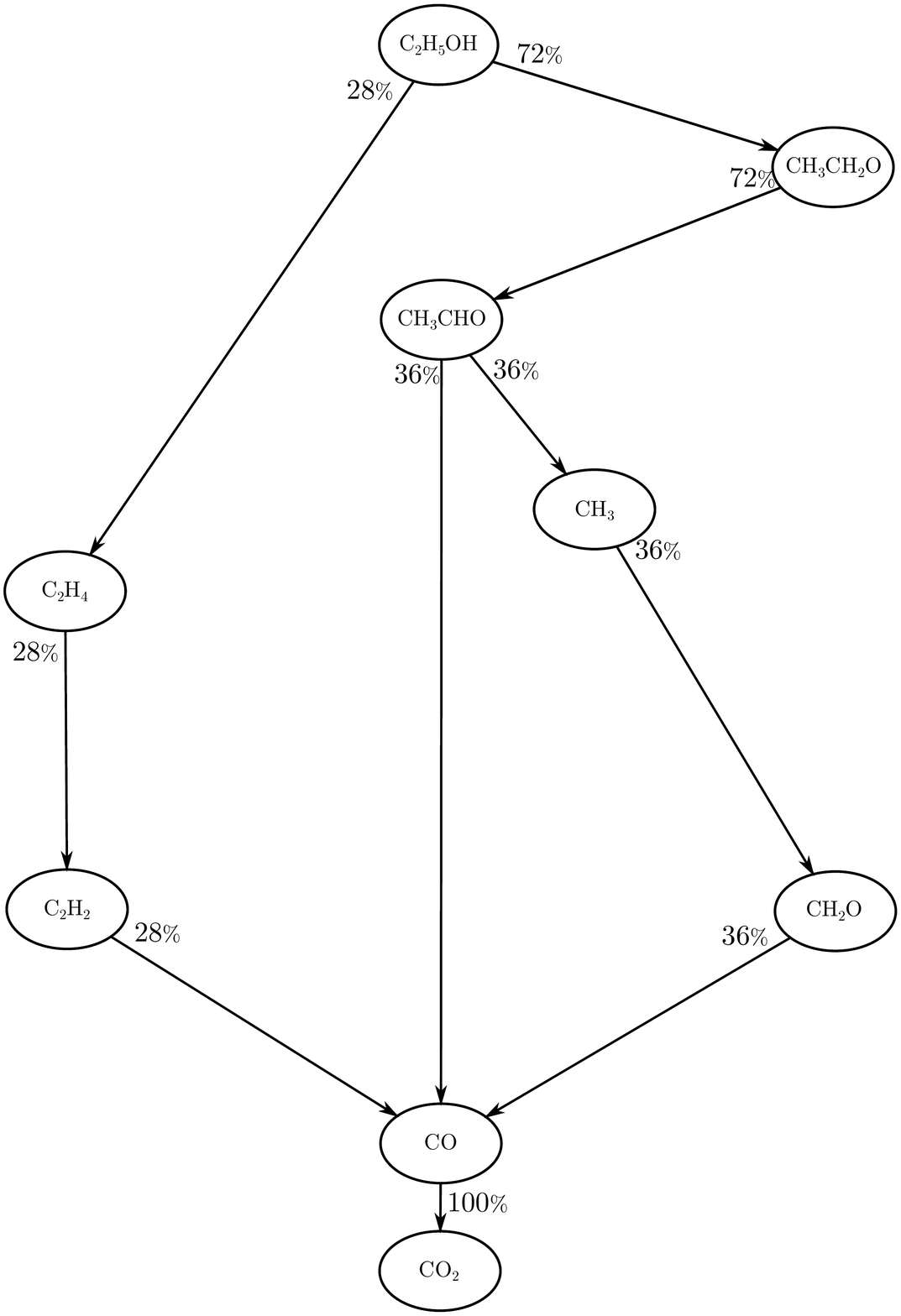} \\
		(a) Skeletal mechanism, $\phi=0.6$ & (b) Reduced mechanism, $\phi=0.6$ \\
		& \\
		\includegraphics[width=.38\textwidth]{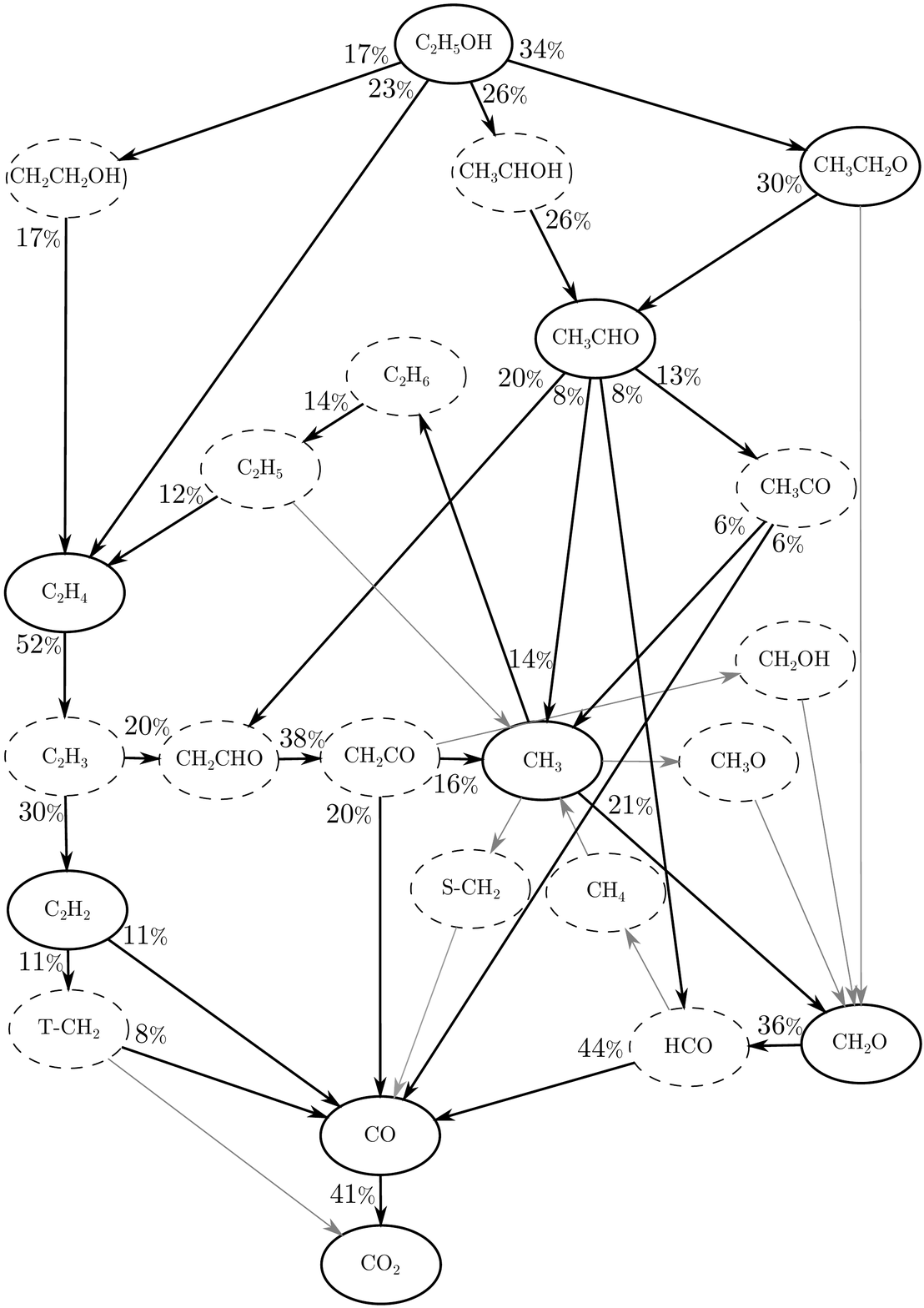} &
		\includegraphics[width=.38\textwidth]{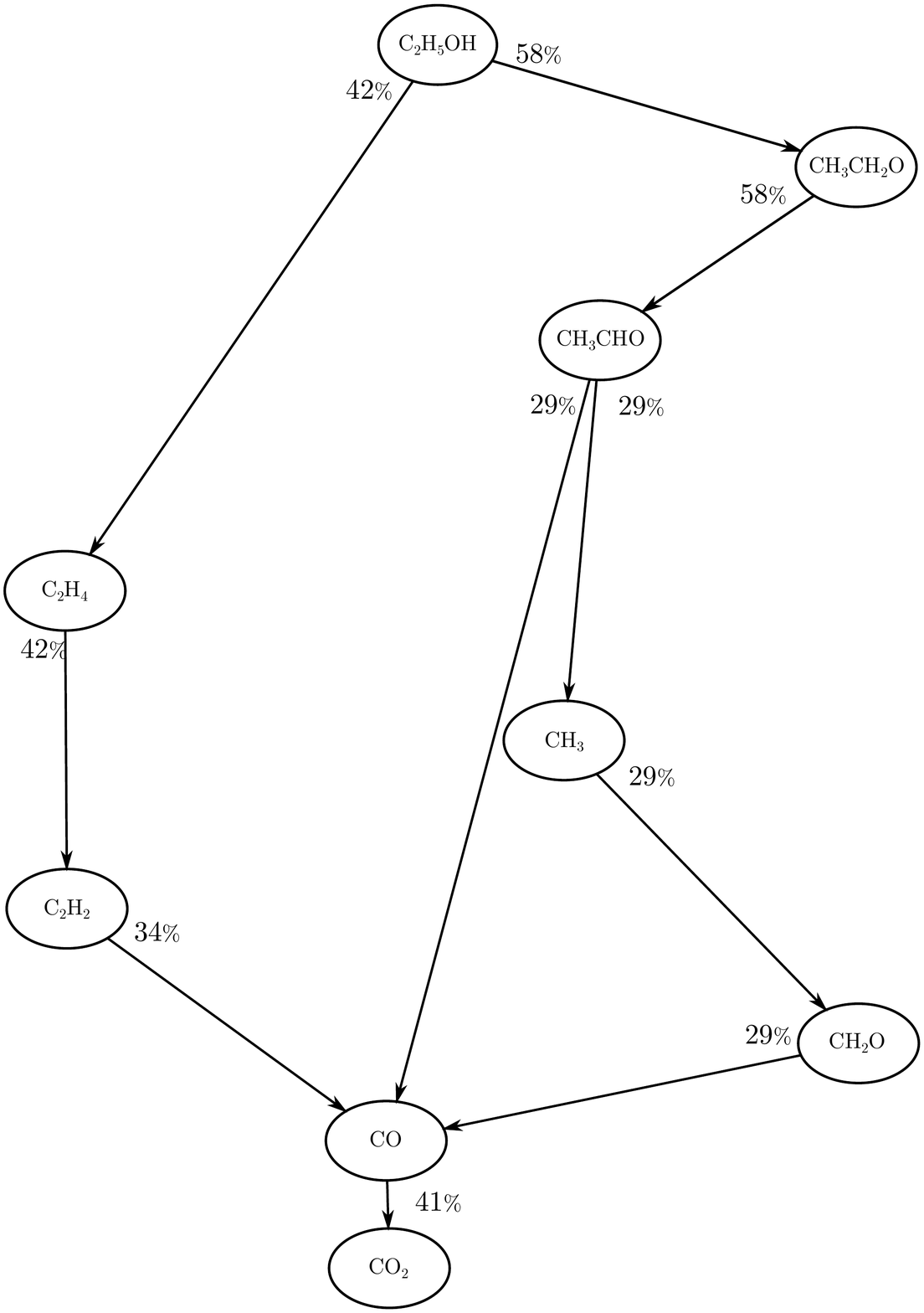} \\
		(c) Skeletal mechanism, $\phi=1.5$ & (d) Reduced mechanism, $\phi=1.5$ \\
	\end{tabular}
	\caption{Reaction-pathway diagrams for ethanol-air premixed flames based on the 66-steps skeletal mechanism and 14-steps reduced mechanism. The upper and lower rows correspond to lean ($\phi=0.6$)  and rich ($\phi=1.5$) flames, respectively. The left and right columns correspond to the skeletal and reduced mechanisms, respectively, at normal atmospheric pressure and an initial temperature of 300 K. The species in steady state are identified by dashed circles in subfigures \emph{a} and \emph{c} on the left.
	}
	\label{fig:pathway_skeletal}
\end{figure}

\noindent
The skeletal mechanism can be further reduced by introducing steady-state assumptions for intermediates, as indicated in the abstract. Extensive computational testing of steady-state approximations for all species was made by calculating the relative fractional differences between the {production and consumption rates $|(\omega_p - \omega_c)/\max(\omega_p, \omega_c)|$}, in the test problems identified above, then requiring the relative fractional differences to be sufficiently small (never reaching 0.1, for example) for introduction of the steady state. Most of the species identified in the abstract are radicals that are expected to be present in small concentrations and would be anticipated to maintain accurate steady states. The radicals \ce{OH} and \ce{CH3} also would be expected to obey accurate steady-state approximations, and indeed they do, according to our criteria.
The concentrations of these last two species, however, appear in the rate expressions for many of the elementary steps, thereby preventing the derivation of an explicit sequential computational procedure, such as that to be given later, if they are eliminated through their steady states. For this reason they are retained as species present in the systematically reduced mechanism, as had to be done previously for \ce{OH} in our development of the systematically reduced mechanism for methanol \cite{FSSW2016}. If steady-state approximations had been introduced for all of the species that maintain accurate steady states, necessitating performing iterations or resorting to special computer programs, then the following 14-step mechanism would be reduced to a 12-step mechanism, having only two fewer species than the 16 to be retained, which could produce only a small advantage, if any, at best. \\
\noindent
Besides applying to radicals, the steady-state approximation is introduced here for three stable species as well, namely \ce{CH4}, \ce{C2H6}, and \ce{CH2CO}. We previously found \cite{FSSW2016} that, in methanol combustion, methane is created in very small concentrations in side reactions and is then efficiently consumed by radicals. Not surprisingly, that was found to remain true for methane in ethanol combustion as well, according to the present computations, and, as may be expected, it also applies to ethane, for the same kinds of reasons that produce this result for methane in methanol flames. Possibly more surprising may be the finding of a steady state for ketene; it arises directly along one of the two H-abstraction paths of acetaldehyde and is significant in adjusting relative formation rates of formaldehyde (more dominant under lean conditions) and methyl through variations of the rates of its two consumption paths, but apparently both of those consumption rates are fast enough to keep its concentration low and to maintain its steady state. \\
\noindent
The overall steps for the reduced mechanism are taken to be
\begin{align}
\ce{H + OH}      & \stackrel{\ce{I}}    {\rightleftharpoons} \ce{H2O}   \label{eq:R_I} \\
\ce{2 H + M}     & \stackrel{\ce{II}}   {\rightleftharpoons} \ce{H2 + M}         \\
\ce{H + HO2}     & \stackrel{\ce{III}}  {\rightleftharpoons} \ce{2 OH}           \\
\ce{H + O2}      & \stackrel{\ce{IV}}   {\rightleftharpoons} \ce{HO2}            \\
\ce{2 OH + M}    & \stackrel{\ce{V}}    {\rightleftharpoons} \ce{H2O2 + M}       \\
\ce{CO + OH}     & \stackrel{\ce{VI}}   {\rightleftharpoons} \ce{CO2 + H}        \\
\ce{CH2O}        & \stackrel{\ce{VII}}  {\rightleftharpoons} \ce{CO + H2}        \\
\ce{CH3 + O2}    & \stackrel{\ce{VIII}} {\rightleftharpoons} \ce{CH2O + OH}      \\
\ce{C2H2 + O2}   & \stackrel{\ce{IX}}   {\rightleftharpoons} \ce{2 CO + H2}      \\
\ce{C2H4 + M}    & \stackrel{\ce{X}}    {\rightleftharpoons} \ce{C2H2 + H2 + M}  \\
\ce{CH3CHO}      & \stackrel{\ce{XI}}   {\rightleftharpoons} \ce{CH3 + CO + H}   \\
\ce{CH3CH2O + M} & \stackrel{\ce{XII}}  {\rightleftharpoons} \ce{CH3CHO + H + M} \\
\ce{C2H5OH + OH} & \stackrel{\ce{XIII}} {\rightleftharpoons} \ce{CH3CH2O + H2O}  \\
\ce{C2H5OH}      &\stackrel{\ce{XIV}}   {\rightleftharpoons} \ce{C2H4 + H2O} \label{eq:R_XIV}     
\end{align}
\noindent
The rates of reactions (\ref{eq:R_I})-(\ref{eq:R_XIV}) and the concentrations of the steady-state species, needed to evaluate the elementary reaction rates, are summarized in appendix \ref{sec:apendice_B}.
~\\
\noindent
The right-hand side of Figs. \ref{fig:pathway_skeletal}b and \ref{fig:pathway_skeletal}d shows the pathways obtained with the reduced mechanism for lean and rich flames. The final form of the 14-steps reduced mechanism was chosen, as far as possible, so that the final reaction pathways correspond to those of the more detailed skeletal mechanism. After the systematic reduction there are only two paths for consumption of the fuel ethanol, the steady state for $\beta$-hydroxyethyl having combined that abstraction with the unimolecular formation of ethylene, and the steady state for $\alpha$-hydroxyethyl having now combined its path with the path through ethoxy (a radical that does not obey an accurate steady state) to acetaldehyde. The simplifications associated with this 14-step mechanism lead to acetaldehyde always producing methyl and carbon monoxide in equal proportions and to all of the methyl passing through formaldehyde. A notable difference for the rich flames is that, after the systematic reduction, ethylene becomes the dominant product species having two carbon atoms, rather {than} the acetaldehyde of the skeletal mechanism, thereby making the systematically reduced mechanism inappropriate for use in detailed investigations of pollutant production in these flames.        
~
\clearpage
\section{Comparisons of predictions}
\begin{figure}[!htp]%
	\begin{tabular}{cc}
		\includegraphics[width=0.47\textwidth]{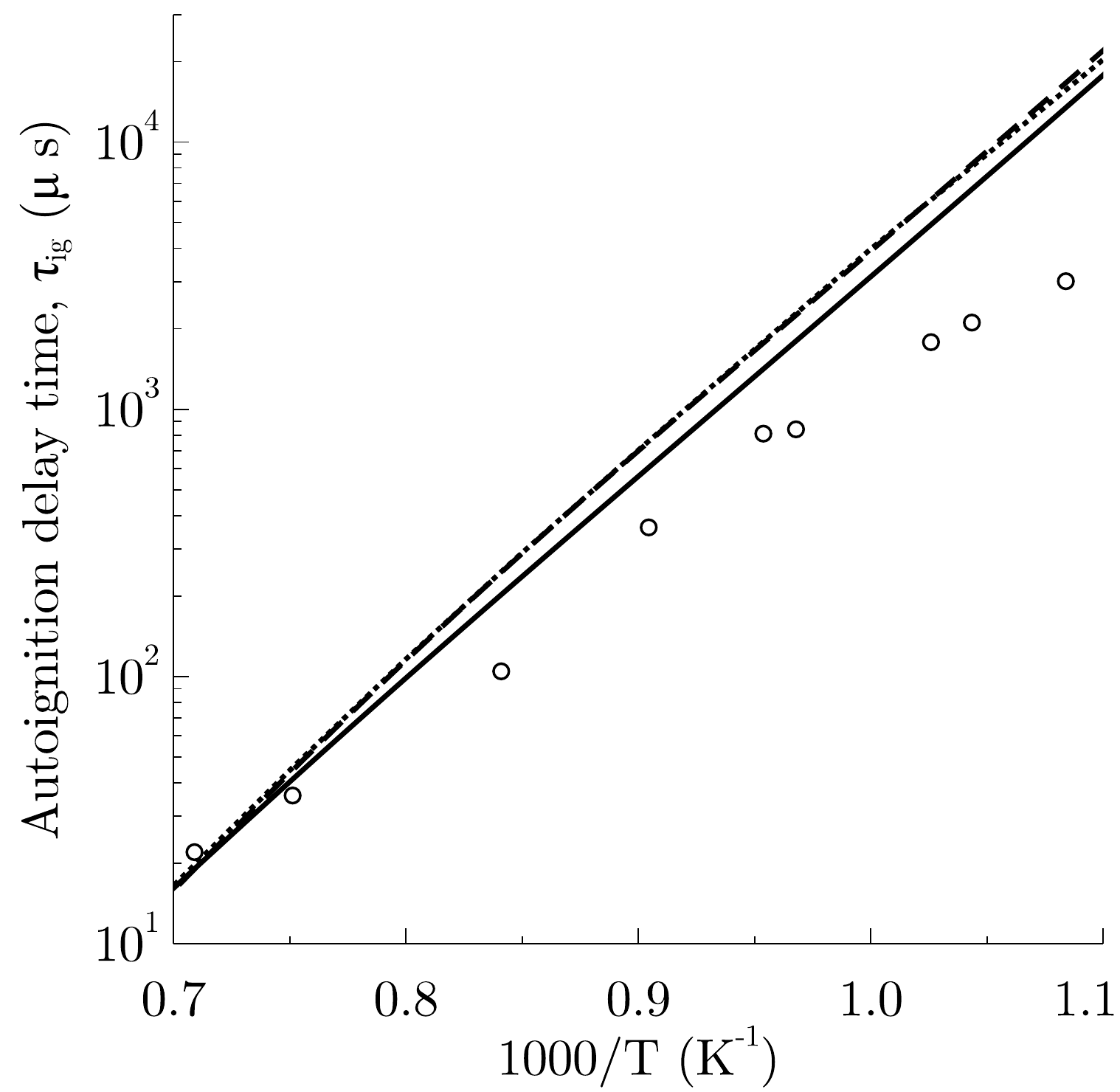} & 
		\includegraphics[width=0.47\textwidth]{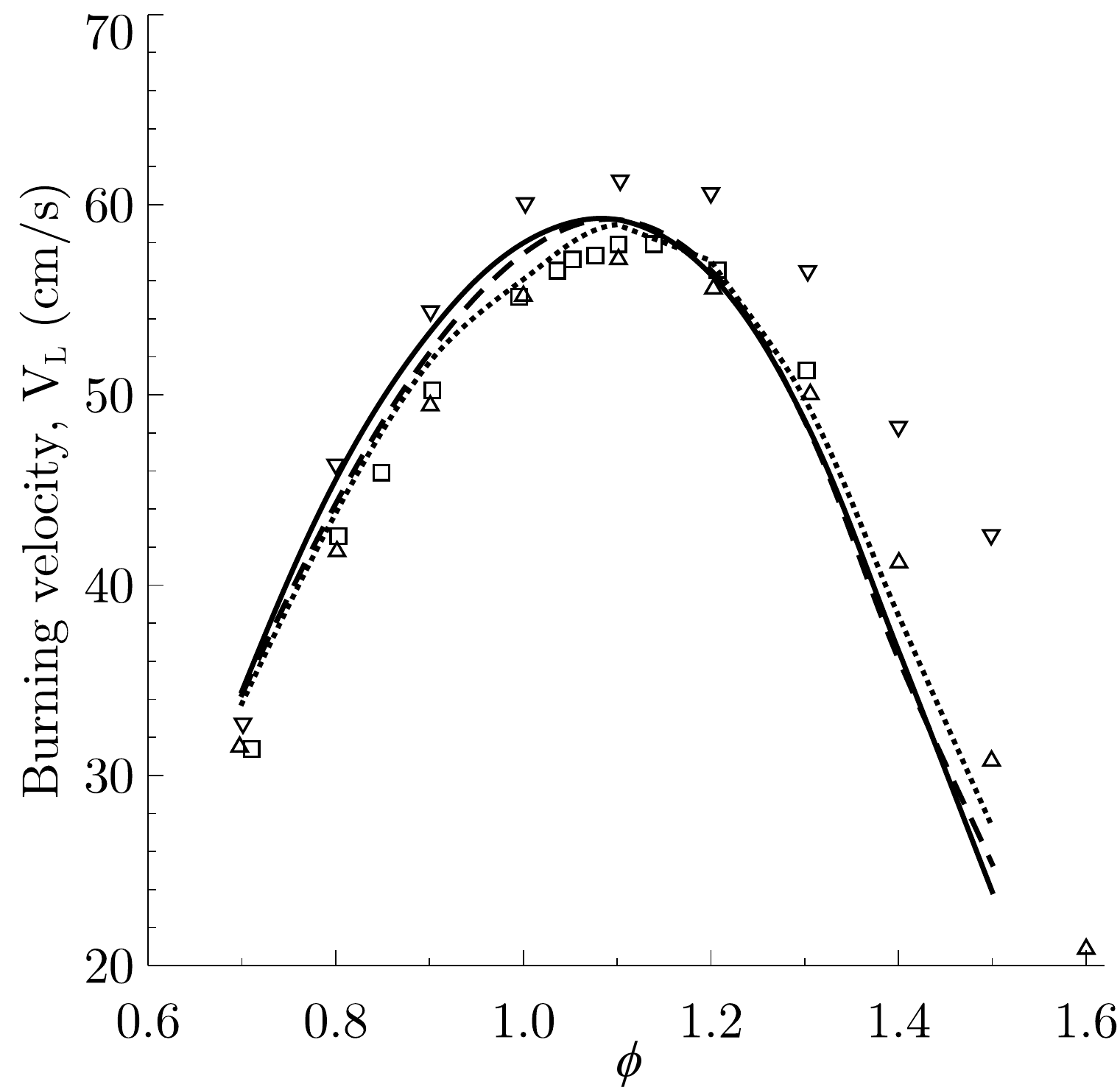} \\
		(a) & (b)
	\end{tabular}
	\caption{Subfigure \emph{a} shows autoignition time for ethanol-air mixtures at $p=13$ bar. 
		Subfigure \emph{b} shows burning velocity for ethanol-air premixed flames as a function of the equivalence ratio $\phi$, at atmospheric pressure for fresh gases temperature $T=373$K.
		The lines represent numerical results, according to 
		{\color{black} \hdashrule[0.5ex][c]{10mm}{2pt}{5mm 0mm} }: San Diego Mechanism, 
		{\color{black} \hdashrule[0.5ex][c]{10mm}{2pt}{3mm 2mm} }: 66-steps Skeletal Mechanism, 
		{\color{black} \hdashrule[0.5ex][c]{10mm}{2pt}{1mm 1mm} }: 14-steps Reduced Mechanism. 
		The symbols correspond to experimental results shown in Fig.~\ref{fig:detailed_chemistry}.    }
	\label{fig:lam_01y02_Res}%
\end{figure}

\begin{figure}[!htp]%
	\begin{tabular}{cc}
		\includegraphics[width=0.47\textwidth]{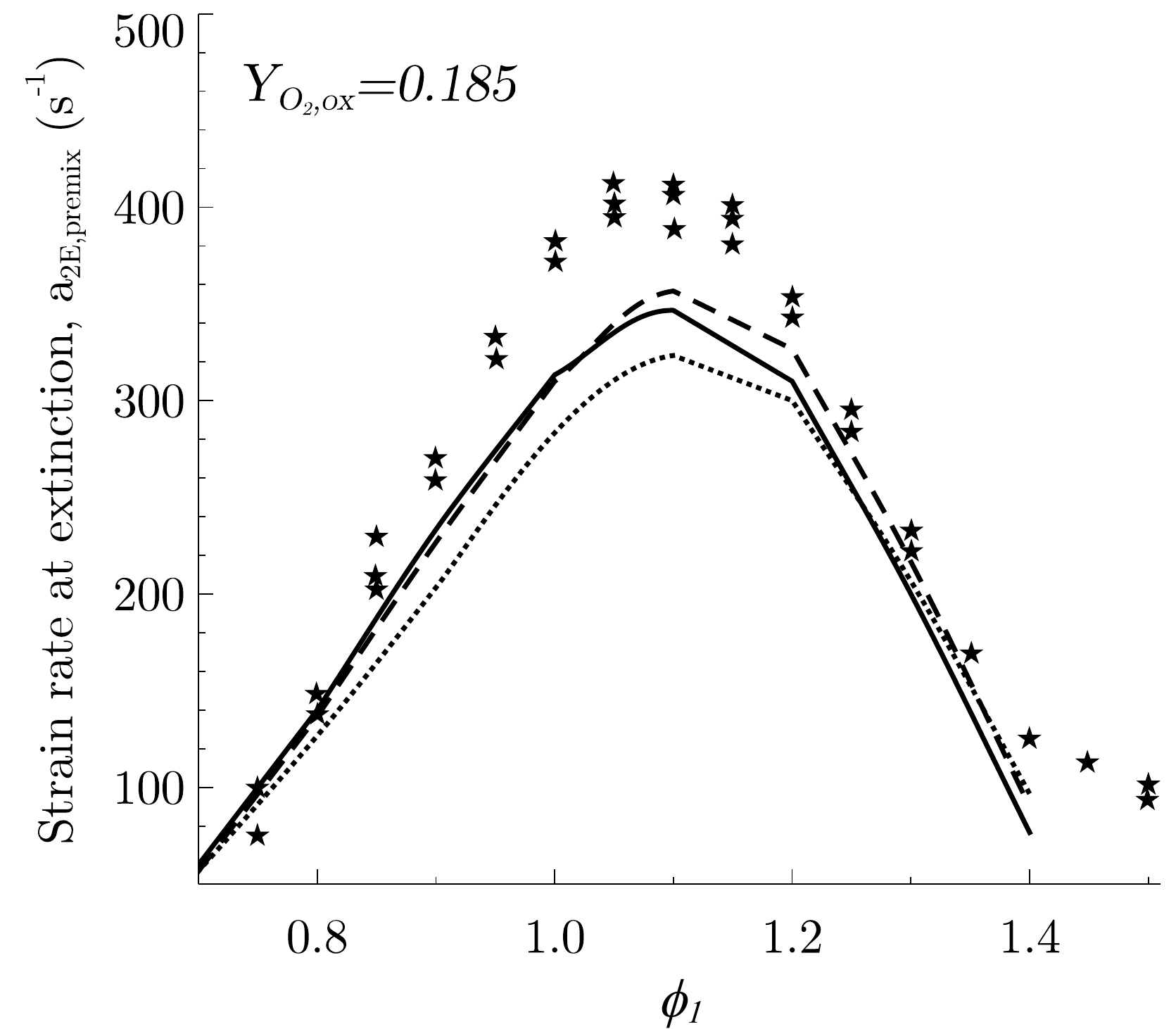}&
		\includegraphics[width=0.47\textwidth]{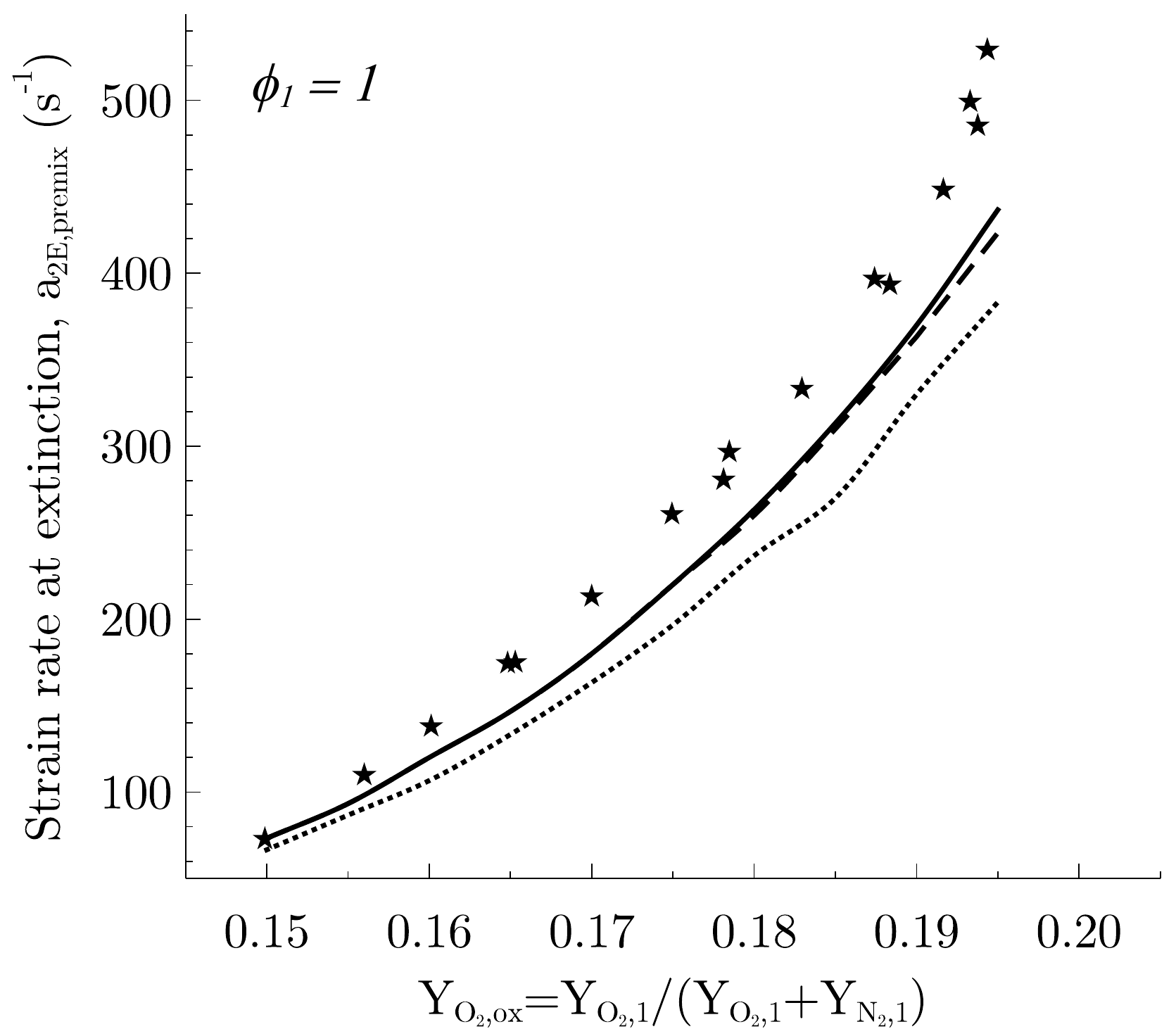}\\
	\end{tabular}
	\caption{Effects of equivalence ratio (left plot) and mass fraction of the oxidizer, $Y_{O_2,ox}=Y_{O_2,1}/(Y_{O_2,1}+Y_{N_2,1})$ (right plot) 
		on the strain rate at extinction for ethanol premixed flame, at 1 atm, for a premixed-gas temperature of T = 323K
		and a nitrogen-stream temperature of T = 298K, for ethanol-oxygen-nitrogen mixtures. The lines represent numerical results
		obtained with detailed mechanism and the symbols correspond to experimental results, detailed as follow:
		{\color{black} \hdashrule[0.5ex][c]{10mm}{2pt}{5mm} }:     San Diego Mechanism
		, {\color{black} \hdashrule[0.5ex][c]{10mm}{2pt}{3mm 3pt} }: 66-steps Skeletal Mechanism
		, {\color{black} \hdashrule[0.5ex][c]{10mm}{2pt}{1mm 1mm} }: 14-steps Reduced Mechanism
		, {\color{black} $\star$}: Seiser et al. \cite{Sei07}.	}
	\label{fig:ext_01y02_Red}%
\end{figure}

\begin{figure}[!htp]%
	\begin{tabular}{cc}
		\includegraphics[width=0.47\textwidth]{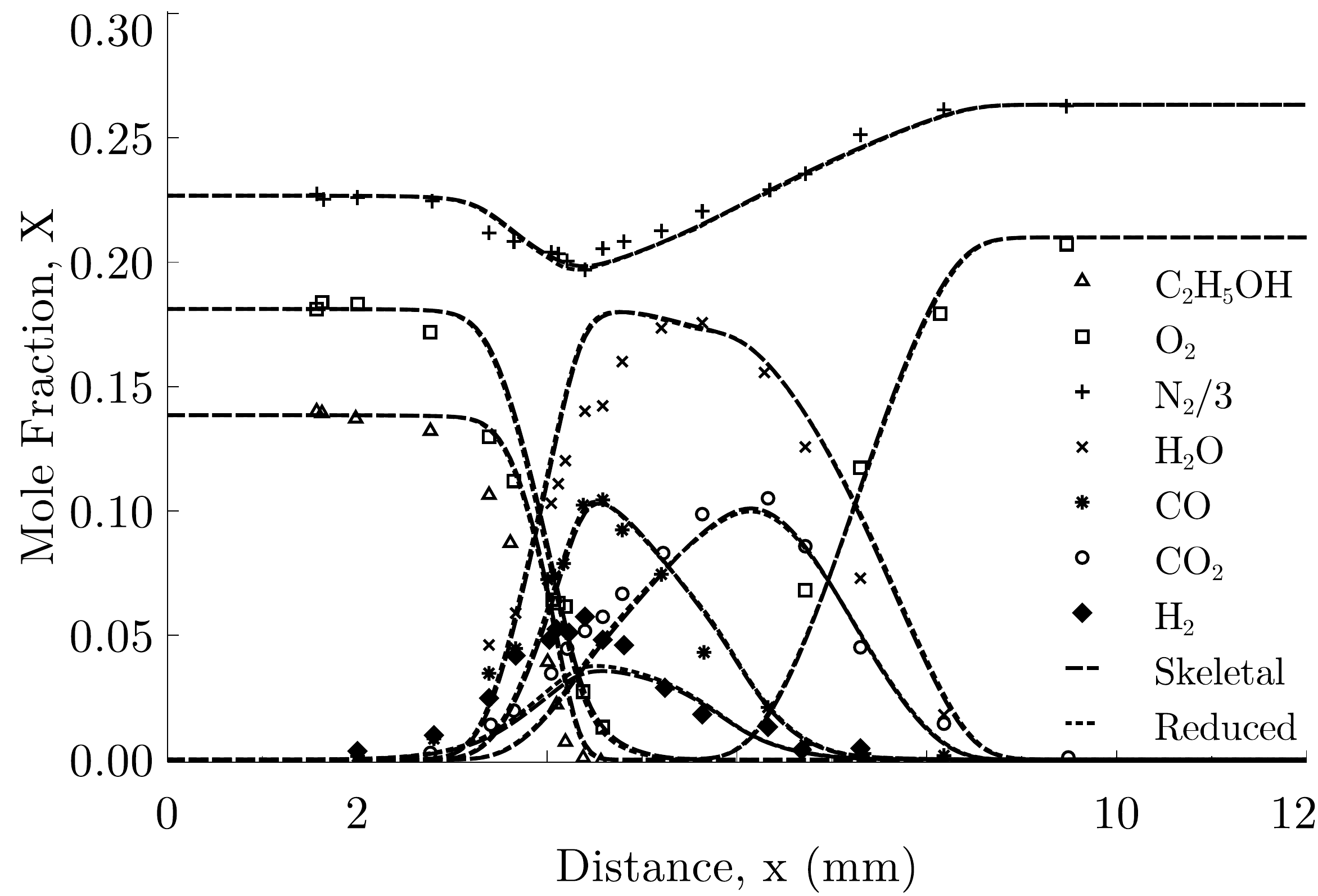}&
		\includegraphics[width=0.47\textwidth]{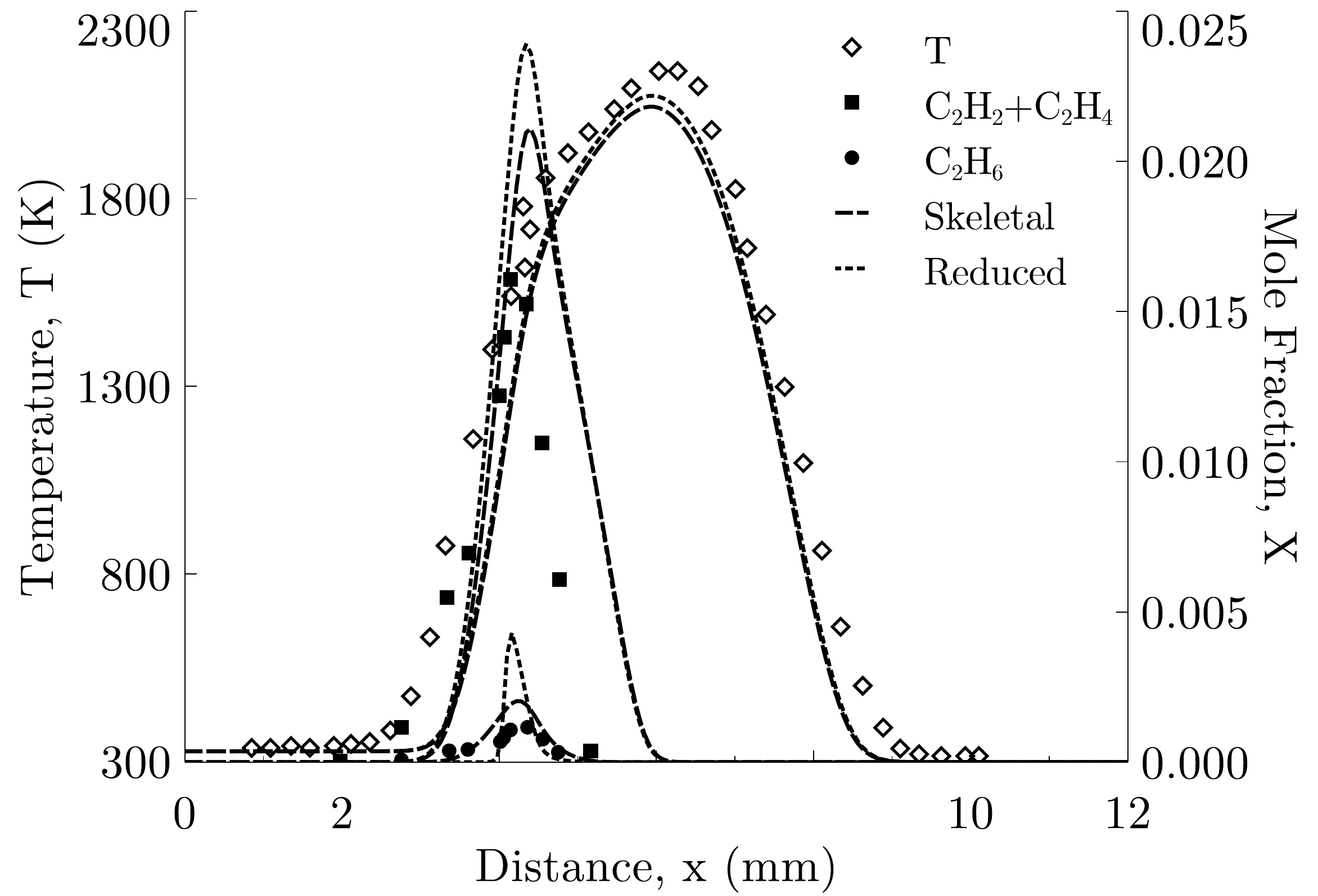}\\
	\end{tabular}
	\caption{Comparisons of predicted and experimental flame structures for partially premixed counterflow flames.
		The lines represent numerical results, according to  
		{\color{black} \hdashrule[0.5ex][c]{10mm}{2pt}{3mm 2mm} }: 66-steps Skeletal Mechanism, 
		{\color{black} \hdashrule[0.5ex][c]{10mm}{2pt}{1mm 1mm} }: 14-steps Reduced Mechanism. 
		The symbols correspond to experimental results of Saxena et al. \cite{Sax07}.}
		\label{fig:prof_01y02_Red}%
	\end{figure}

\noindent
The quantitative tests shown in table \ref{tab:01} indicate errors in both the skeletal and reduced mechanisms that are quite comparable with those of the detailed mechanism for the problems addressed (differences from burning-velocity and autoignition-time measurements being well below 10 \%),  while they both afford gains in computational times for premixed-flame problems. Those gains are likely to be larger for more complex calculations, such as those needed in studying turbulent combustion.\\
\noindent
Predictions of the detailed, skeletal, and reduced mechanisms for the experiments of Fig. \ref{fig:detailed_chemistry} are compared in Fig. \ref{fig:lam_01y02_Res}, where it is seen that the differences between these predictions are quite small for these cases, as was found also for other autoignition and burning-velocity tests. Predictions of burning velocities at elevated pressures by the reduced mechanism, in fact, fortuitously are in almost exact agreement with experiment \cite{Bro13}, while the detailed and skeletal mechanisms exhibit small overpredictions. Computations of stoichiometric flame structures at normal atmospheric conditions showed that predictions of peak concentrations by the skeletal and reduced mechanisms for major stable species typically differed by less than 10 \%, never exceeding about 30 \%, while for some of the steady-state radicals the differences ranged from a factor or 2 to a factor of 5, possibly affecting some of the chemical pathways comparably. Differences for the stable side-produced species methane and ethane, however, often were found to be much greater than that, thereby indicating possible substantial inaccuracies in predictions of the concentrations of these two species by the reduced mechanism; while it is worth keeping this observation in mind, these differences are unlikely to affect any other conclusions appreciably, since those species do not play any role in other chemical pathways.\\
\noindent
Comparisons of predictions and experiments \cite{Sei07} for counterflow premixed-flame extinction conditions in flows against a cold nitrogen stream are shown in Fig. \ref{fig:ext_01y02_Red}. Here strain rates at extinction are plotted as functions of the equivalence ratio for an oxygen mass fraction in the oxygen-nitrogen oxidizer stream added to the fuel stream of 0.185 (on the left) and for variations of that oxygen mass fraction in stoichiometric mixtures (on the right). As was previously observed for counterflow diffusion flames \cite{Sax07}, the detailed San Diego mechanism tends to underpredict the experimentally observed extinction strain rates. In the plot on the left it is seen that the predictions of the skeletal mechanism lie very close to those of the detailed mechanism but are slightly closer to the experimental data for rich flames, while the plot on the right indicates virtually no difference between these two predictions. Although not very different from the other predictions, the results of the reduced mechanism show lower strain rates at extinction than the skeletal mechanism, noticeably farther from the experimental data. Along with the comparisons shown on the right in Fig. \ref{fig:lam_01y02_Res}, these results are representative of the accuracies to be expected in premixed-flame computations when using the simplifications developed herein.\\
\noindent
Extinction conditions for diffusion flames were calculated with the skeletal and reduced mechanisms and were found to differ very little from those obtained with the detailed mechanism. The differences between the predictions of extinction conditions for the different mechanisms for diffusion flames are comparable with or less than those seen in Fig. \ref{fig:ext_01y02_Red} and therefore are not included here. Skeletal-mechanism calculations of diffusion-flame extinction with mixture-average and multicomponent transport descriptions using CHEMKIN \cite{Chem} exhibited the same substantial differences as those reported previously \cite{Sax07} for the detailed mechanism.\\
\noindent
Reduced-mechanism predictions of measured counterflow diffusion-flame structures were previously reported for methanol \cite{FSSW2016}, and predictions of the detailed San Diego mechanism were reported and discussed previously for corresponding counterflow ethanol flames \cite{Sax07}, exhibiting generally good agreement with experiment. Counterflow ethanol diffusion flames are very thin, leading to difficulties in gas sampling that translate to inaccuracies in measured temperature and concentration profiles, but partial premixing of oxidizer into the fuel stream significantly spreads out the counterflow flame, improving measurement accuracy, while providing essentially the same tests of chemical-kinetic predictions. Comparisons of predictions of the detailed mechanism with experiments for these partially premixed flames were made previously \cite{Sax07}, yielding good agreements, and Fig. \ref{fig:prof_01y02_Red} shows corresponding comparisons for the skeletal and reduced mechanisms. The calculations, made with mixture-average transport descriptions like those available in Cosilab \cite{Cos}, were compared with calculations made with CHEMKIN \cite{Chem} employing multicomponent diffusion instead, and in this case (contrary to what is sometimes found for diffusion-flames, especially in extinction) there were negligible differences in the results obtained. The profiles of temperature and of concentrations of major species predicted by the skeletal mechanism, shown in Fig. \ref{fig:prof_01y02_Red}, are indistinguishable from the profiles \cite{Sax07} found with the detailed chemistry, and they are seen in the figure to be nearly identical to those obtained with the reduced mechanism.\\
\noindent
For the minor stable species, \ce{C2H2}+\ce{C2H4}, \ce{C2H6} and \ce{CH4} (the last not shown in Fig. \ref{fig:prof_01y02_Red}), the predictions of the skeletal mechanism are somewhat different from those of the detailed mechanism, resulting in more of the first two (around 20 \% more \ce{C2H2}+\ce{C2C4} and nearly twice as much \ce{C2H6}, found by comparing the results shown here with those reported previously \cite{Sax07} but not plotted here) and predicting much less \ce{CH4} (less than 1/3 as much as is shown in the previous reference). In addition, the reduced mechanism is seen in the figure to give noticeably more \ce{C2H2}+\ce{C2H4} than the skeletal mechanism, and a narrower and taller (by a factor of 2) \ce{C2H6} peak, much higher than the measurement. These observations further emphasize substantial differences in predictions of the skeletal and reduced mechanism for minor stable species concentrations and also demonstrate inaccuracies in predictions of the skeletal mechanism for those species. It may be concluded in general that, apart from inaccurate predictions for these minor stable species and specially questionable usefulness of the reduced mechanism for investigations of pollutant production, both the skeletal and reduced mechanisms perform quite well. \\    
\\
\\
\clearpage
\section{Conclusions}
\noindent
The need for accurate computations requires using more detailed models for chemistry, molecular transport and thermodynamics, which  leads to costly  computations in terms of CPU time and memory. According to Hilbert et al \cite{Hilbert2004}, up to 80-90 \% of the total computing time is usually spent in calculating diffusion velocities and chemical source terms. With the objective of reducing the computational time while retaining accuracy of the predictions, a reduced-chemistry description that involves only 14 overall steps among 16 reactive chemical species has been derived by introducing chemical-kinetic steady-state approximations for the relevant intermediate species, using, as starting point, a skeletal mechanism of 66 steps and 31 chemical species, also developed in this work.\\
\noindent
The skeletal and reduced chemistry description described above achieved computation time savings of 70 \% and 80 \%, respectively, when compared with the computational time of the detailed mechanism. Also, the lower number of chemical species of the reduced mechanism  will contribute to shorten the computation of the diffusion velocities. In general, the more accurate the transport model is, the greater the time reduction becomes. The number of species is especially critical for detailed transport models, such as the multicomponent transport model \cite{Dixon1968}, because it determines the dimension of the matrix that needs to be inverted in order to obtain the molecular transport terms. \\
\noindent
Greater computational savings can be obtained by further specializing the reduced mechanism. In simpler flames, hydrogen peroxide \ce{H2O2} and  hydroperoxyl \ce{HO2} radicals are found to be in steady state \cite{Fer09, Pet93}. In ethanol flames, additional intermediate radicals are in steady state ($\ce{CH3CH2O}$, $\ce{CH3CHO}$, $\ce{CH3}$ and $\ce{CH2O}$), suggesting a clear path to develop shorter and faster reduced mechanism. A similar approach can be used if a more reduced mechanism specific for autoignition is needed; in this case \ce{H2O2} and \ce{HO2} cannot be assumed to be in steady state, but simplifications come from other species being in steady state, as was found in the case of methanol \cite{Seiser2011}.\\
\noindent
As shown in previous sections, both the skeletal and reduced mechanisms provide accurate predictions for flame structure and global properties of ethanol--air combustion (auto-ignition times, flame propagation and even extinction of both premixed and non-premixed strained flames). A driving force in the development of reduced mechanisms is the optimization of combustion systems which nowadays cannot be separated from emissions reductions objectives, as dictated by pollutant abatement policies. Some of the pollutants (such as \ce{CO}) can be predicted directly by the reduced mechanism described above. Others, (like \ce{NO_x}), can be obtained by post-processing the flame structure using an specific mechanism \cite{Hill_Smoot_2000}. Predictions of other pollutants, such as soot, would rely on the capability to obtain accurate profiles of PAH precursors, such as \ce{CH4}, \ce{C2H2}, \ce{C2H4} and \ce{C2H6} \cite{Wang_1997, Blanquart_2009}. In this regard, the skeletal mechanism, which exhibits inaccuracies in predictions of these precursors and does not even include other precursors, such as smaller ring compounds, requires augmentation to assess soot emissions, an area of great interest for biofuels use \cite{Nature_2017}.
\section{Acknowledgements}
\noindent
This work was supported by Project ENE2015-65852-C2-1R of the Spanish \textit{Ministerio de Econom\'{\i}a y Competitividad}.


\end{document}